\documentclass[runningheads]{llncs}

\usepackage[T1]{fontenc}
\usepackage[english]{babel}
\usepackage{amsmath}
\usepackage{mathtools}
\usepackage{bm}
\usepackage{algorithm}
\usepackage[noend]{algpseudocode}
\usepackage{xargs} 
\usepackage{xcolor}
\usepackage{soul}
\usepackage{wrapfig}
\usepackage{graphicx}
\usepackage{hyperref}
\usepackage{comment}
\usepackage{tabularx}
\usepackage{booktabs}
\usepackage{multicol}
\usepackage{multirow}
\usepackage{cite}
\usepackage{subfigure}

\hypersetup{
    colorlinks=true,
    linkcolor=blue,
    filecolor=magenta,      
    urlcolor=cyan,
    citecolor=cyan,
}

\newcommand{\GG}[0]{\textsc{GraphGuess}}

\usepackage{tikz}
\newcommand{\ballnumber}[1]{\tikz[baseline=(myanchor.base)] \node[circle,fill=.,inner sep=1pt] (myanchor) {\color{-.}\bfseries\footnotesize #1};}

\graphicspath{{figures/}}

\title{\textsc{GraphGuess}: Approximate Graph Processing System with Adaptive Correction}
\titlerunning{\textsc{GraphGuess}}

\author{
    Morteza Ramezani
    \and
    Mahmut T. Kandemir
    \and
    Anand Sivasubramaniam
    }
\authorrunning{M. Ramezani et al.}
\institute{
    \vspace{-5pt}
    Pennsylvania State University\\
    \email{\{morteza,kandemir,anand\}@cse.psu.edu}
}

\begin{document}

\maketitle

\vspace{-10pt}
\begin{abstract}
Graph-based data structures have drawn great attention in recent years. The large and rapidly growing trend on developing graph processing systems focuses mostly on improving the performance by preprocessing the input graph and modifying its layout. These systems usually take several hours to days to complete processing a single graph on high-end machines, let alone the overhead of pre-processing which most of the time can be dominant. Yet for most graph applications the exact answer is not always crucial, and providing a rough estimate of the final result is adequate. Approximate computing is introduced to trade off accuracy of results for computation or energy savings that could not be achieved by conventional techniques alone. In this work, we design, implement and evaluate GraphGuess, inspired from the domain of approximate graph theory and extend it to a general, practical graph processing system. GraphGuess is essentially an approximate graph processing technique with adaptive correction, which can be implemented on top of any graph processing system. We build a vertex-centric processing system based on GraphGuess, where it allows the user to trade off accuracy for better performance. Our experimental studies show that using GraphGuess can significantly reduce the processing time for large scale graphs while maintaining high accuracy.
\vspace{-5pt}
\end{abstract}
\section{Introduction}
Nowadays graph-based data are pervasive, with applications including  search engines, social and biological networks and financial systems. 
Studies~\cite{yuhanna2013techradar} have pointed out that graph-based data structures constitute more than $25\%$ of all enterprise data.
Graph sizes are increasing rapidly and they consist of billions of nodes and edges with relatively random patterns, posing significant challenges to computer systems and architecture.
Hence, efficient batch processing or serving instantaneous interactive queries on these graphs becomes a challenge in the big data era. 
Observing this need, several graph processing frameworks~\cite{kyrola2012graphchi, low2014graphlab, nai2015graphbig} have been introduced to reduce the programming burden and avoid the need for extensive optimizations for each and every application.

Most prior graph processing systems try to find  the ``exact answer" in a resource-efficient and timely manner. 
However, in many real world applications (especially in the large-scale data analytics domain), the exact answer may not be necessary all the time, and one can usually tolerate some amount of error. 
For example, web search engines are most often interested only in the first few tens or hundreds of pages of what users are looking for, disregarding the rest~\cite{mcsherry2015scalability, mitliagkas2015frogwild}.
Also, in a financial security application where the goal is to find fraudulent activity patterns, it is good enough to just capture a rough estimate of the number of times the pattern  occurs~\cite{iyer2018asap}. 
Such characteristics of many modern graph applications can allow the system to trade off accuracy for execution efficiency.  
Motivated by this, in this paper, we pursue approximate computing techniques for a spectrum of graph applications where we strive to provide faster and more efficient output with high quality results.

The fundamental question in approximation is the relationship between the amount of data processed and the associated accuracy of the final results. 
Based on this relationship, typically, the solutions either run the algorithm on a portion of data (sampling) or run a part of the algorithm (task skipping, interpolation) on the entire data, to achieve a reasonable approximation of what actual results would be. 
Unlike the other types of data structures, the randomness of graphs makes it difficult to exploit specific properties in the data that may help isolate them for conventional sampling and/or subsetting of the processing.

While the system side of graph processing community has been focusing on running the exact algorithms~\cite{kyrola2012graphchi, malewicz2010pregel, low2014graphlab, ching2014giraph}, the theoretical side has come up with a large body of graph approximation techniques, which try to provide a mathematical bound for the solution. 
Yet such approaches can impose a huge burden on the programmer to design and implement new and complex graph algorithms. 
There exist few prior works aiming at practical aspects when approximating graph application executions \cite{shang2014auto, iyer2018bridging, besta2019slim, iyer2018asap, heidarshenas2020v}.
However, such existing approaches not only rely on offline preprocessing, which imposes a huge overhead and is unfeasible in many cases where the graph structure changes rapidly, but also are not general and mainly target a limited type of applications.

Motivated by these limitations of prior approaches and ever-growing importance of graph applications, in this work, we propose \GG{}, a run-time adaptive approximation model for graph processing systems which: (i) requires minimal preprocessing and change to the original graph and applications to figure out what data to include in the computation; (ii) adapts dynamically to the graph structure and application at hand; (iii) preserves characteristics of the original graph and increases the output accuracy; and (iv) significantly reduces the volume of computation performed compared to the exact graph computation. 
Although our evaluations are confined to static graphs in this paper, \GG{} is certainly applicable to dynamic graphs as well.

\section{Graph Processing Systems}
\label{section:processing}

\subsection{Think Like a Vertex}

Traditionally, processing a large graph required significant developer efforts to design and implement an optimized version of the algorithm. 
Upon increasing interest in the applications with underlying graph-based data structures, this task has become increasingly more challenging and inefficient. Several general-purpose graph processing frameworks have been introduced and evaluated in recent years to improve the programmability of graph applications, focuses mainly on performance and/or scalability. These frameworks include, but not limited to, vertex-centric \cite{malewicz2010pregel, ching2014giraph, low2014graphlab}, edge-centric \cite{roy2013x}, data-centric \cite{nguyen2013lightweight}, matrix operation based \cite{sundaram2015graphmat} and task based \cite{kulkarni2007optimistic}.
Among all these models, the idea of ``\emph{think like a vertex}'' or \emph{vertex-centric} programming model has seen significant interest and widespread deployment in recent works \cite{low2014graphlab, gonzalez2012powergraph, kyrola2012graphchi}, and is our model of choice.

\begin{wrapfigure}[10]{R}{0.625\textwidth}
\vspace{-35pt}
\begin{minipage}{1\linewidth}
\begin{algorithm}[H]
 \caption{An example of vertex-centric algorithm with Gather-Apply-Scatter (GAS) model}
 \label{alg:vc}
 \begin{algorithmic}[1]
    % \small
    \For{each Vertex \texttt{v}}
        \For {each incoming Edge \texttt{e}}
            % \State {\texttt{temp} $\gets$ \textsc{Gather}\texttt{(e)}} \;
            \State {\texttt{new\_property} $\gets$ \textsc{Gather}\texttt{(e)}} \;
        \EndFor
        \State {\texttt{property} $\gets$ \textsc{Apply}\texttt{(new\_property)}} \;
        \For {each outgoing Edge \texttt{e}}
            \State {\textsc{Scatter}\texttt{(e)}} \;
        \EndFor
    \EndFor
\end{algorithmic}
\end{algorithm}
\end{minipage}
\end{wrapfigure}
Vertex-centric model is an iterative approach that executes a so-called \emph{vertex program} that includes one or more user-defined functions (\emph{udf}) for each vertex in every iteration. 
To eliminate the overhead of unnecessary computations in vertices that have not seen any updates, the concept of \emph{active vertices} is employed, where a list of vertices is maintained to keep the vertices that have received an update in the previous iteration for sub-setting the processing (i.e., reducing its scope in the next iteration).
{\em Gather-Apply-Scatter} (GAS) \cite{low2014graphlab} is one of the widely used vertex-centric models, which consists of three main phases that are executed in each iteration. 
First, in the \emph{Gather} phase, a vertex reads from all incoming edges and reduces them to a single value.
Next, in the \emph{Apply} phase, a vertex uses the reduced value to compute its own property. 
Lastly, in the \emph{Scatter} phase, each vertex propagates its new value over all out-going edges. 
A sample program in a vertex-centric model is shown in Algorithm \ref{alg:vc}.

\subsection{Preprocessing the Graph}
\label{subsection:preproc}

While vertex-centric models provide ease of programming, their performance can still throttle as the graph size increases. 
However, due to the random nature of the graph, as the number of edges increases, the system suffers from poor spatial locality when accessing the vertex properties. 
Also, a processed (destination) vertex is unlikely to be processed again before most of the other vertices (low temporal locality). 
Hence, the system performance still suffers from an under-performing cache, and cannot benefit from any known prefetching techniques~\cite{ahn2016scalable}. 
Finally, real world graphs are known to follow the power-law distribution and have skewed degree nodes, which makes synchronization in shared-memory systems problematic~\cite{capelli2018ipregel}. 
As a result, many existing frameworks focus on optimizing the data layout, by preprocessing the graph and reordering the vertices and/or edges.
Preprocessing the graph can improve the locality of the input and increase the performance at run time. 
However, altering and rewriting the original graph requires several iterations over the entire graph, which may exceed the actual running time of the application itself, making these methods impractical \cite{mukkara2018exploiting}.

\subsection{Approximate Analysis}

Approximate computing has gained much popularity in big data processing systems in recent years.
Several techniques are used to run the application approximately including, running the application on a smaller portion of the dataset (``sampling") or running the program partially on the entire data (``task skipping"). 
Note that, in either case, there is an underlying assumption that data are independent and the accuracy will improve (linearly) with more data or more tasks processed. 
The same holds true for the area of hardware-based approximation \cite{jevdjic2017approximate} as well, where data representation is approximated (e.g., via quantization) in favor of performance, bandwidth, storage, or power gains. 

\begin{wrapfigure}[17]{r}{0.5\textwidth}
    \vspace{-24pt}
    \centering
    \subfigure[]
    {
      \includegraphics[width=0.80\linewidth]{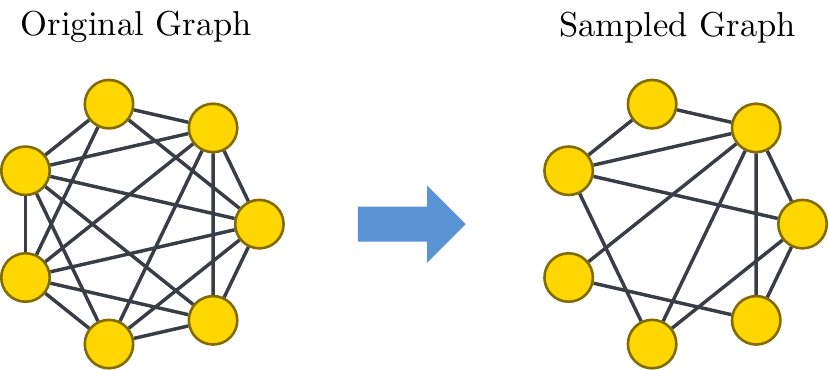}
      \label{fig:random-sampling}
    }
    \vspace{-10pt}
    \subfigure[]
    {
        \includegraphics[width=0.71\linewidth]{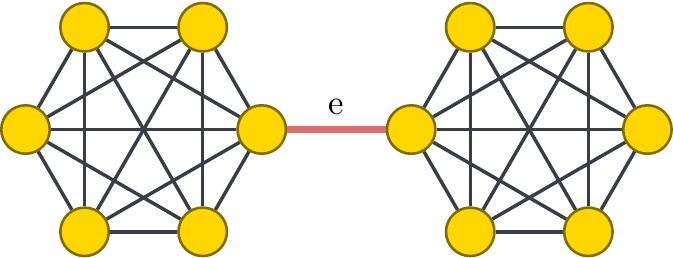}
        \label{fig:sample-problem}
    }
    \vspace{-5pt}
    \caption{
    (a) \emph{Sparsifying} with uniform sampling with $\sigma=0.5$.
    (b) Example of Dumbbell, where random uniform sampling may end up not choosing $e$.
    }
    \vspace{-10pt}
\end{wrapfigure}
Reducing the size of the input graph or skipping part of the process are promising solutions for certain problems in graph processing.
However, unlike other types of data structures, there is a dependency between data elements (vertices) in the graph, where an error in a single vertex (as a result of approximation, for example) can propagate to the entire application. 
There have been several works in the theory community on graph approximation algorithms~\cite{qiao2012approximate, spielman2011spectral, mitliagkas2015frogwild, abraham2016fully, bernstein2020fully}.
In most cases, the proposed approximation algorithms are variants of the corresponding original algorithms, with proper changes to reduce their running times, while bounding the error through randomization. 
However, since such approaches are very ``algorithm-specific", they are not readily applicable in existing general graph processing systems that are used today to run a wide variety of applications.

Generating a smaller graph or \emph{graph summarization} techniques are introduced to speed up graph processing. Among these methods, sampling, similar to approximate analysis, is one key idea in theoretical graph approximation, commonly referred to as \emph{graph sparsification}.
In this approach, a set of edges (or vertices) are selected randomly from the original graph, to reduce the amount required processing.
A parameter determines the degree of sparsification, and the accuracy of the result depends on this parameter. An example of random uniform sampling is illustrated in Figure~\ref{fig:random-sampling}.
Apart from sparsification, other graph summarization techniques such as graph sketching~\cite{sarlos2006randomize}
and graph compression \cite{shin2019sweg, besta2019slim} have been proposed in recent years. In addition, a few graph approximation frameworks have been developed to alleviate the performance bottleneck of large graph processing~\cite{iyer2018asap, shang2014auto, iyer2018bridging}. 
Shang et al.~\cite{shang2014auto} propose an automatic approximation for graph computing, which targets compiler-level optimization rather than runtime system optimization. 
ASAP~\cite{iyer2018asap} targets approximate graph pattern mining (finding pattern in graphs).
More recently, V-Combiner~\cite{heidarshenas2020v} is proposed with a similar goal as graph sketching techniques. 
V-Combiner consists of an initial step to create an approximate graph with fewer vertices (and edges) and a delta graph to use during the recovery phase in order to compute the output for missing vertices.
Similar to the previous methods, this technique depends on building a couple of new graphs, which means a large memory burden in addition to the preprocessing overhead.

%%%%%%%%%%%%%%%%%%%%%%%%%%%%%%%%%%%%%%%%%%%%%%%%%%%%%%%%%%%%%%%%%%%%%%
\subsection{When Graph Approximation Fails}
\label{subsec:graph-approx}

While sparsification reduces the graph size, the reduced graph may {\em not} necessarily preserve all essential properties of the original graph which are critical for the target application. 
One such problem can be seen in Figure~\ref{fig:sample-problem}, which is referred to as the ``dumbbell graph". 
In this case, uniform sampling can omit edge $e$ which attaches the two parts of the graph, leading to serious errors in the algorithms that rely on this edge (e.g., graph connectivity). 
To make sure that edge $e$ is chosen, one may need to sample several  times, which is usually quite inefficient and makes random uniform sampling error prone. 
Spielman et. al.~\cite{spielman2011spectral} proposed a sparsification technique based on the degree of the nodes.
However, this method may also fail in some scenarios~\cite{spielman2011graph}. 

% \begin{wrapfigure}[9]{r}{0.65\textwidth}
\begin{figure}
    \centering
    \vspace{-10pt}
    \includegraphics[width=0.75\textwidth]{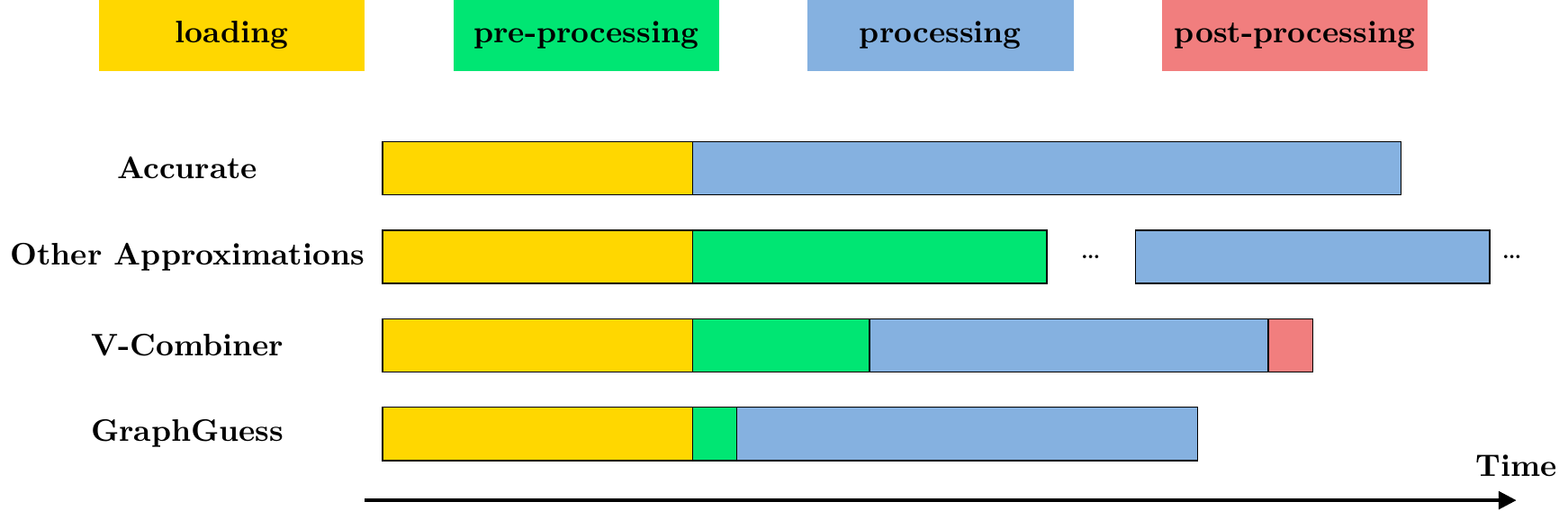}
    \vspace{-5pt}
    \caption{General timeline of executing a program on graph processing systems.}
    \label{fig:new-design}
    \vspace{-10pt}
\end{figure}
% \end{wrapfigure}

To solve the problem of leaving out very important edges when sampling, Spielman et. al. \cite{spielman2011graph} use importance for edges or ``effective resistance" that is taken into account when sparsifying the graph. 
In a graph, the effective resistance of an edge $e$ is equal to the probability that the edge $e$ appears in a random spanning tree. 
Although a quite powerful metric, computing effective resistance for {\em all} edges in a given large graph can introduce excessive overheads at preprocessing time~\cite{heidarshenas2020v, spielman2011spectral}.

In Figure~\ref{fig:new-design}, we show the timeline of running several graph approximation techniques for an algorithm on a relatively large graph (we scale the figure for better presentation).
Clearly, graph summarization and V-Combiner are still suffering from the same type of problem, where the additional overhead is justifiable if the algorithm runs for a large number of iterations. 
In this paper, inspired by the existing works in graph theory, we introduce \GG{}, which requires minimal preprocessing, with no need for a new graph. The performance benefits of \GG{} come mainly from reduced number of edges, which has been shown as a main factor in slowing down the graph processing systems~\cite{shang2014auto, iyer2018bridging, iyer2018asap,heidarshenas2020v}, while its higher accuracy is due to adaptive correction.
Furthermore, \GG{} provides flexibility that allows integration with all types of graph algorithms and requires minimal changes to the front-end applications.

%%%%%

\section{GraphGuess}
\label{section:graphguess}

\subsection{Programming Model}
\label{subsection:pm}

The programming model used in \GG{} is closely similar to that in the vertex-centric model discussed earlier in Section~\ref{section:processing}. Such a design makes our framework significantly easier to adapt to the existing applications. Here, for the sake of simplicity, we use \emph{pull-based} vertex-centric model. Note however that \GG{} is not limited to any specific underlying model  and can be easily adapted to the others. The functions defined in the \GG{} programming interface and their descriptions are provided below, alongside an example of PageRank implementation using this interface shown in Algorithm~\ref{alg:pr-vc}.
\begin{itemize}
    \setlength\itemsep{0em}
    \item \textsc{GG-Gather}: Gathers property from incoming edges in each iteration and computes a local function.
    The red line here is the minimal change required in user-program introduced by \GG{}. 
    \item \textsc{GG-Apply}: Applies the newly calculated property to a given vertex.
    \item \textsc{GG-VStatus}: Checks the convergence criteria based on the old and new values and activate the vertex. 
\end{itemize}

\subsection{Tracking the Edge Influence}

% % \begin{wrapfigure}[11]{R}{0.55\textwidth}
% % \vspace{-42pt}
% % \begin{minipage}{0.55\textwidth}
% \begin{algorithm}[H]
%  \setlength{\intextsep}{0pt} 
%  \caption{PageRank using \GG{} API} 
%  \label{alg:pr-vc}
%  \scriptsize
%  \begin{algorithmic}[1]
%     \Statex
%     \Procedure{gg-gather}{vertex \texttt{u}, vertex \texttt{v}}
%         \State \texttt{old\_value $\gets$ u.property}
%         \State \texttt{u.property} $\gets$ $\frac{\texttt{v.property}}{\texttt{v.degree}}$ $+$ \texttt{u.property}
%       \textcolor{red}{\State \Return $\frac{\texttt{u.property - old\_value}}{\texttt{u.property}}$} \Comment{\textcolor{red}{\small{Return the \emph{Edge Influence}}}}
%     \EndProcedure
%     \Statex ...
%     \Statex Other functions (which are unchanged compared to normal VC) are eliminated due to space constraint
%     \Statex ...
%     \end{algorithmic}
% \end{algorithm}
% % \end{minipage}
% % \end{wrapfigure}

\begin{algorithm}[ht]
 \caption{PageRank using \GG{} API} 
 \label{alg:pr-vc}
 \small
 \begin{algorithmic}[1]
    \Statex
    \Procedure{gg-gather}{vertex \texttt{u}, vertex \texttt{v}}
        \State \texttt{old\_value $\gets$ u.property}
        \State \texttt{u.property} $\gets$ $\frac{\texttt{v.property}}{\texttt{v.degree}}$ $+$ \texttt{u.property}
      \textcolor{red}{\State \Return $\frac{\texttt{u.property - old\_value}}{\texttt{u.property}}$} \Comment{\textcolor{red}{\small{Return the \emph{Edge Influence}}}}
    \EndProcedure
    \Statex 
    \Procedure{gg-apply}{vertex \texttt{u}, value \texttt{property}}
        \State \texttt{u.old\_property} $\gets$ \texttt{u.property}
        \State \texttt{u.property} $\gets$ $\frac{(1 - \texttt{d})}{\texttt{N}} + \texttt{d}\times{}\texttt{property}$
    \EndProcedure
    \Statex
    
    \Procedure{gg-vstatus}{vertex \texttt{u}}
        \If{$\left|\texttt{u.old\_property} - \texttt{u.property}\right| > \epsilon$} 
        \texttt{u.active} $\gets$ \texttt{True}
        \Else{}
        \texttt{u.active} $\gets$ \texttt{False}
        \EndIf
    \EndProcedure
    \Statex
    
    \Procedure{gg-estatus}{edge \texttt{e}, value \texttt{influence}, value \texttt{threshold}}
        \If{\texttt{influence} $>$ \texttt{threshold}}
        \texttt{e.active} $\gets$ \texttt{True}
        \Else{}
        \texttt{e.active} $\gets$ \texttt{False}
        \EndIf
    \EndProcedure
    \end{algorithmic}
\end{algorithm}

\begin{figure}
  \vspace{-15pt}
  \centering
  \subfigure[PageRank]{
    \includegraphics[width=0.45\linewidth, scale=1.5]{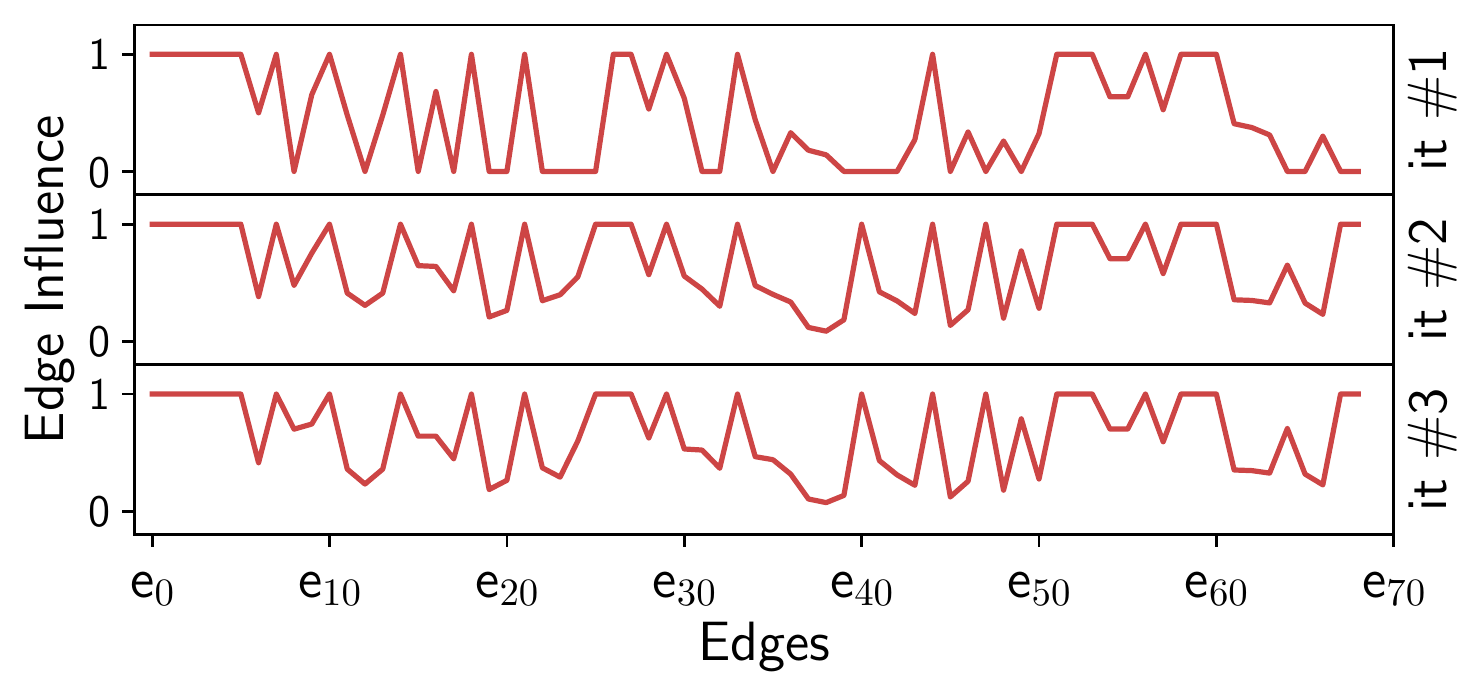}
    \label{fig:pr-changes}
  }
  \subfigure[SSSP]{
  \vspace{-10pt}
    \includegraphics[width=0.45\linewidth, scale=1.5]{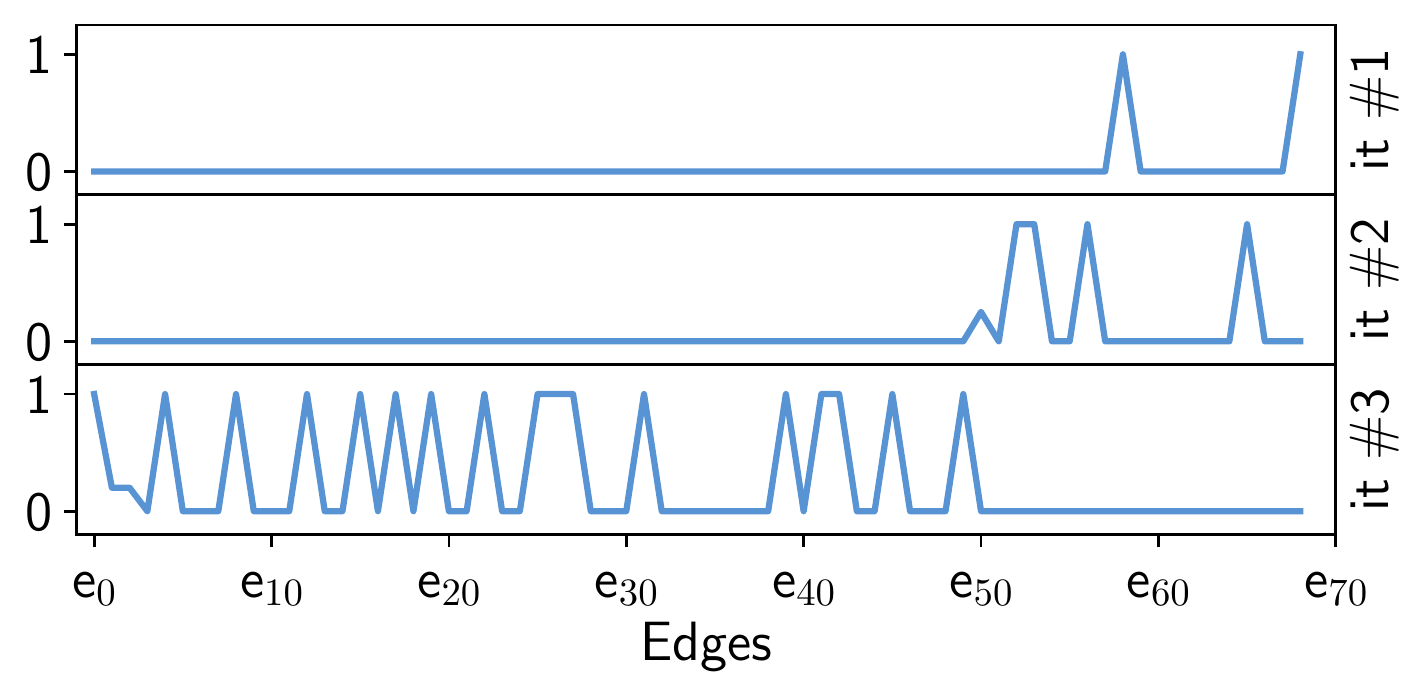}
    \label{fig:sssp-changes}
  }
  \vspace{-13pt}
  \caption{The edge influence for 3 iterations on (a) PageRank and (b) SSSP}
  \vspace{-5pt}
\end{figure}

To avoid the high preprocessing overheads, \GG{} tries to find the importance of edges dynamically at runtime using the concept of \emph{edge influence}. 
In each iteration, the calculations at a given vertex are impacted by the ``influence" of the edges, i.e., a function of the property of other end vertices. If one could track this importance (\emph{edge influence}), it would be possible to dynamically figure out whether that edge should continue to be used in subsequent iterations as well. 
This is also a natural way to exploit user-defined functions that get computed at each vertex, rather than artificially try to identify the importance of the edges offline. 
In other words, regardless of the type of the algorithm (distance, value propagation, etc.) \GG{} would be able to automatically extract the \emph{edge influence} information. 
To achieve this flexibility, one can slightly modify the \texttt{GG-Gather()} function in the vertex-centric model from Algorithm \ref{alg:pr-vc} to capture the influence of each edge (Line 4). 
This additional information can be passed through to the main method in the processing system, and be used for the future iterations, details of which will be discussed later in Section~\ref{subsection:add}.

Figure~\ref{fig:pr-changes} illustrates the \emph{edge influence} for all edges in a small, synthetically-generated graph running PageRank algorithm for 3 iterations (one figure for each iteration), using the modified \texttt{GG-Gather()} in Algorithm \ref{alg:pr-vc}.
As shown, those edges which provide higher influence would continue to have higher impact in future iterations as well. Thus, one can eliminate the edges that are not contributing significantly to the final result, to reduce the number of processed edges within each iteration.  This technique has been previously proposed in the work by McSherry \cite{mcsherry2015scalability} as an optimization ``\emph{solely}" for the PageRank algorithm.
While PageRank showed a relatively non-changing edge influences across iterations, that may not necessarily be the case in other applications.
For instance, Figure~\ref{fig:sssp-changes} shows the edge influences for the SSSP algorithm in which {\em not} all vertices are active all the time, as it is a traversal algorithm.  Consequently, the \emph{edge influence} values depend not only on their source and destination, but also on the iteration. As mentioned before, the criteria for determining ``importance" will itself vary, in general, across applications. 
As a result, a single criteria may not suffice for tracking edge importance, and it is much more practical to let the user determine the criteria as part of the programming exercise.

%%%%%%%%%%%%%%%%%%%%%%%%%%%%%%%%%%%%%%%%%%%%%%%%%%%%%%%%%%%%%%%%%%%%%%
\subsection{Runtime Modes}
\label{subsection:add}

While the vertex (user) program for \GG{} requires minor changes, the underlying processing system in \GG{}, which is completely oblivious to the developers, still needs a few modifications to accommodate the approximate computing capability. One of our main goals in \GG{} is to avoid any unnecessary preprocessing and building new graphs. Hence, we include a {\em flag} for each edge in the graph to determine if the edge is active or not.  
Next, we define the following two running modes for \GG{}.
\vspace{-5pt}
\begin{itemize}
    \item \textbf{Accurate Mode:} This is the default setting, where each vertex reads data from all incoming edges (regardless of their active flag) and executes the corresponding functions.
    \item \textbf{Approximate Mode:} Each vertex only reads and processes data from its "active" incoming edges and disregards the rest.
\end{itemize}
\vspace{-5pt}

In the \emph{approximate mode}, each vertex deals with fewer edges, thereby reducing the total number of processed edges and less processing time. 
However, this also means that vertices do not have access to the \emph{edge influence} values for those inactive edges.
To figure out whether those missing edges continue to be immaterial in the computation, we define the concept of a \emph{superstep}, in which the system switches back to the \emph{accurate mode}, enabling each vertex to pull information from {\em all} its incoming edges.

\subsection{Adaptive Correction}
\label{subsection:adaptive}

In \GG{}, the system starts in the \emph{approximate mode} where a subset of edges is deactivated, similar to graph sparsification, discussed in Section~\ref{subsec:graph-approx}. 
A control parameter, $\bm\sigma$ (\emph{sparsification parameter}) controls the number of active edges, with a higher value of $\bm\sigma$ indicating more active edges. 
We want to emphasize that the overhead of this part is negligible and in fact it can be done while loading the graph into the system.
The system continues in the \emph{approximate mode} for $\bm\alpha$ iterations (\emph{approximate window}), and then performs a \emph{superstep}, where it transitions to the \emph{accurate mode} to adaptively correct the initial edges selection. 
At the end of a \emph{superstep}, \GG{} can determine and activate new ``qualified" edges, based on the computed \emph{edge influence} and a threshold $\bm\theta$ (\emph{influence threshold}).
This is done in a user defined function, \texttt{GG-EStatus}\texttt{()}, example of which is shown in Algorithm \ref{alg:pr-vc}.
The process continues in the next iteration with all activated edges.

\begin{wrapfigure}[9]{r}{0.55\textwidth}
  \vspace{-25pt}
  \centering
  \includegraphics[width=\linewidth]{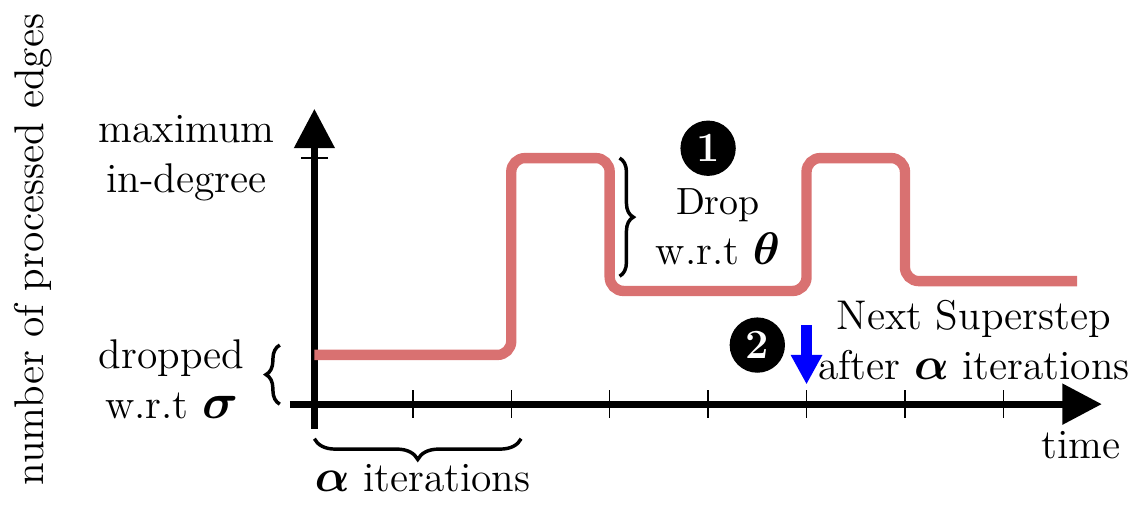}
  \vspace*{-20pt}
  \caption{Time-line of running \GG{}}
  \label{fig:gg-times}
\end{wrapfigure}

This approach can be seen as a coarse-grain active list technique introduced in the original vertex-centric model. 
However, here, in addition to activating vertices based on their property change, for each vertex, its edges are activated based on the \emph{edge influence} computed in the superstep. 
Furthermore, we drop the edges with minimal influence to reduce the number of processed edges in the system.
We only pick the edges that meet the influence threshold $(\bm\theta)$, and discard the rest. 
That is, after performing a superstep, \GG{} only activates the qualified edges and deactivate the rest, regardless of whether they were active before or not.  
Figure~\ref{fig:gg-times} illustrates the Vertex Point of View (\emph{VPV}) evolution time-line of this approach, by showing the number of edges processed across iterations for a single vertex.

In a large graph, performing a single superstep iteration may not suffice to capture the changes that gradually ripple through the graph.
Motivated by this observation, we propose to use periodic supersteps in \GG{} and control the frequency using the same approximate window parameter. More specifically, $\bm\alpha$ controls how long it takes before another superstep should take place, as shown in Figure~\ref{fig:gg-times} \ballnumber{2}. Note that, a smaller value of $\bm\alpha$ means more frequent superstep executions and results in better accuracies; however, it also imposes higher overheads on the system.
While one could have different parameters for controlling the first superstep and their recurrence, and vary them through the course of running the algorithm, we find that using the same parameters provides a good enough accuracy-performance trade-off.

%%%%%
\section{Applications and Error Criteria}
\label{section:application}

\subsection{Applications and Datasets}

As discussed before, \GG{} is application domain agnostic and can work with any target application.
Most current graph benchmark suites include popular graph algorithms, including graph traversal, property computing, and pattern mining. 
Based on that and due to limited space, we selected Single Source Shortest Path \textbf{(SSSP)}, Weakly Connected Components \textbf{(WCC)} Page Rank \textbf{(PR)}, and Belief Propagation \textbf{(BP)}.
We also chose a wide variety of graph workloads: Wikipedia \textbf{(WP)}, LiveJournal \textbf{(LJ)}, Twitter \textbf{(TW)} and Friendster \textbf{(FS)}.

\begin{figure*}[h]
    \centering
    \vspace*{-10pt}
    \includegraphics[width=1\textwidth, angle=0]{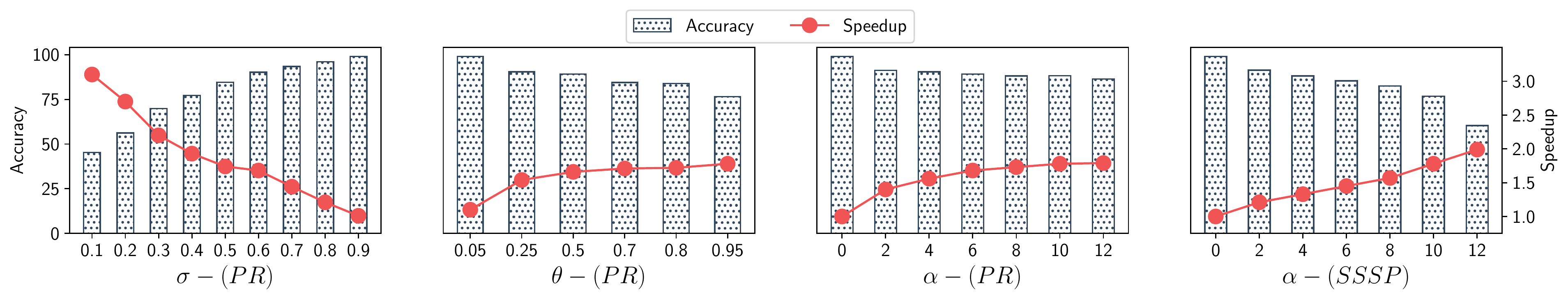}
    \vspace*{-20pt}
    \caption{The impact of \GG{} parameters on the accuracy and speedup of the system running on Wikipedia. 
    The left y-axis shows the accuracy and the right y-axis shows the speedup and the x-axis represents the value of the parameter.}
    \label{fig:param}
\end{figure*}

\subsection{Error Metrics}
\label{subsection:errors}

Unlike non-graph based approximations where determining the error is relatively straightforward, defining error metrics in approximated graph applications is not trivial. Consider PageRank for instance, where the algorithm itself includes a control knob for error (convergence rate), which can also be used for evaluating the approximated output. 
At the same time, since the intention of PageRank is to relatively rank the pages, absolute values for the rank may not matter too much. 
Consequently, in the following we explore different evaluation metrics.

$\bullet$ \textbf{Top-K Error:} (used for PR and BP)
Similar to \cite{mitliagkas2015frogwild} this metric is defined as the fraction of vertices in the top-k ranks of approximated output that are not in the top-k ranks of the accurate output.
    
$\bullet$ \textbf{Relative Error:} (used for WCC)
The relative error is the ratio of the difference between the accurate and the approximate value to the accurate value.

$\bullet$ \textbf{Stretch Error:} (used SSSP)
Borrowed from graph theory, the stretch factor is applicable in most distance-based graph applications and defined as the ratio of the approximated value to the accurate value for each vertex. 
% Note that this metric is suitable for SSSP. To determine the error of approximation, we randomly select $0.1\%$ of the vertices as source and run SSSP to compute the average stretch error.

For the sake of better representation, in all of our evaluations we use accuracy, ranging from $0\%$ to $100\%$, which is defined as $(1 - \text{error})\times 100$.

%%%%%
\section{Experimental Evaluations}
\label{section:results}

\textbf{Implementation and Setup:}
We implemented \GG{} on top of a vertex-centric graph processing platform in C++ and parallelized using the shared memory API OpenMP.
Note that, in principle, \GG{} can be integrated into any other graph processing system (including those with a pre-processing mechanism), with minimal changes in the API.
We use the accurate execution as the {\em baseline}, and measure the speedup and accuracy compared to this baseline.
For a fair comparison, we also implemented the user functions for all of our benchmark applications for \GG{} and all other baselines. 
In all our experiments, unless otherwise stated, we run the experiment for the same number of iterations five times and report the mean value for the metrics. 
To determine the effective performance of \GG{}, we only measure the execution time of processing and pre-processing parts of the application.

%%%%%%%%%%%%%%%%%%%%%%%%%%%%%%%%%%%%%%%%%%%%%%%%%%%%%%%%%%%%%%%%%%%%%%
\subsection{Sensitivity to Control Parameters}
\label{subsec:sense}

To examine the impact of the control parameters on the overall system efficiency and accuracy of \GG{}, we conduct several experiments and present the results in Figures~\ref{fig:param}. 
These figures capture the relationship between different values of control parameters (x-axis) and speedup on the right y-axis (red line) and accuracy on the left y-axis (blue bars) compared to the accurate baseline. 
In each setting, we fix all other control parameters and selectively vary the desired parameter (indicated on the bottom of each figure) to observe its effect. 
We ran the applications on \emph{Wikipedia} dataset for a fixed number iterations to get a fair comparison. 
Due to space constraints, we show the results only for PageRank and SSSP in one instance.

Figure~\ref{fig:param}($\sigma$) plots the speedup-accuracy comparison, for different values of the sparsification parameter ($\bm\sigma$). 
The value of $\bm\sigma$ is varied from $0.1$ to $0.9$ in $0.1$ increments, where $0$ represents the case where no edge is active and $1$ represents the case where all edges are active. 
Clearly, one can expect better accuracy from the system when $\bm\sigma$ has a higher value, but at the same time the performance would be degraded. 
Figure~\ref{fig:param}($\sigma$) confirms this expectation, and with $\sigma = 0.1$ the accuracy is about $40\%$ with speedup up to $3\times$, whereas with a higher value $\sigma = 0.9$, the accuracy increases significantly while the performance improvement ending up being as good as the accurate mode.

The influence threshold or $\bm\theta$, can also affect both accuracy and speedup, as demonstrated in Figure~\ref{fig:param}($\theta$). 
More specifically, a lower value of $\bm\theta$ makes it easier for an edge to get activated -- as a result the system processes more edges and achieves higher accuracy. 
On the other hand, with higher threshold values, the edges are only being activated if they make {\em significant influence} at the supersteps, thus, the system in this case {\em trades off} accuracy for performance. 
Figure~\ref{fig:param}($\theta$) also reveals that, when $\bm\theta$ is changed from $0.05$ to $0.5$ both performance and accuracy change significantly, whereas a change of threshold from $0.5$ to $0.8$ has a relatively low impact on the performance, while resulting in less accuracy, due to limited number of processed edges. 

Figure~\ref{fig:param}($\alpha-PR$) plots the impact of changing the value of $\bm\alpha$ (superstep frequency) for the PageRank algorithm. 
Earlier in Figure~\ref{fig:pr-changes}, we saw that for PageRank, the distribution of edge influence is more or less uniform throughout the execution of the algorithm. 
As a result, the specific starting point (or frequency) for a superstep does not affect the accuracy in any significant way. 
It can be seen here that changing $\bm\alpha$ does not affect the accuracy considerably, compared to the previous results, while doing so can change the speedup. 
However, this is not always the case, especially when the algorithm traverses the graph and vertices get activated later. % as the program progresses.

\begin{wrapfigure}[11]{r}{0.35\textwidth}
    \centering
    \vspace*{-25pt}
    \includegraphics[trim={0 0 9.5cm 0},clip, width=\linewidth]{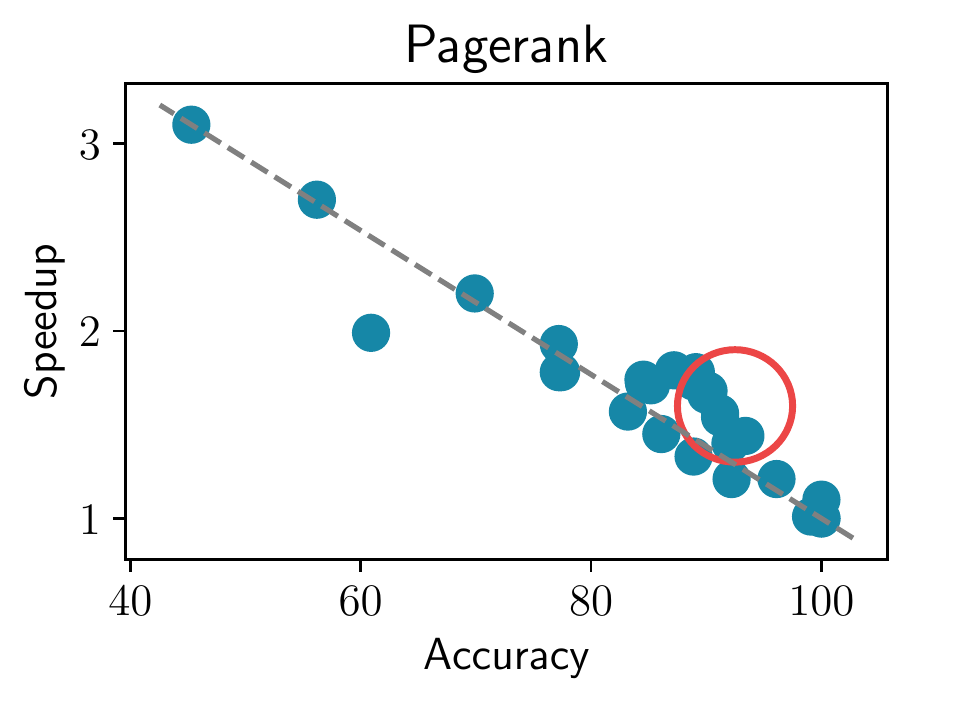}
    \vspace*{-20pt}
    \caption{Accumulated parameter analysis to determine fair value for control parameters.}
    \label{fig:agg-params}
\end{wrapfigure}
An example of such behavior is in the SSSP algorithm, the results of which are given in Figure~\ref{fig:param}($\alpha-SSSP$).
Previously, Figure \ref{fig:sssp-changes} showed that, in this application, the edge influences do not follow the same pattern throughout the time; hence, the time at which a superstep is performed matters more than the PageRank case.  
It can be concluded from this figure that the different values of $\bm\alpha$ impact both error and performance considerably.
In general, to handle all types of application behaviors in \GG{}, we use the combination of control parameters $\bm\sigma$, $\bm\theta$, and $\bm\alpha$. 

\begin{figure*}[ht]
    \centering
    \vspace*{-10pt}
    \includegraphics[width=1.0\linewidth,angle=0]{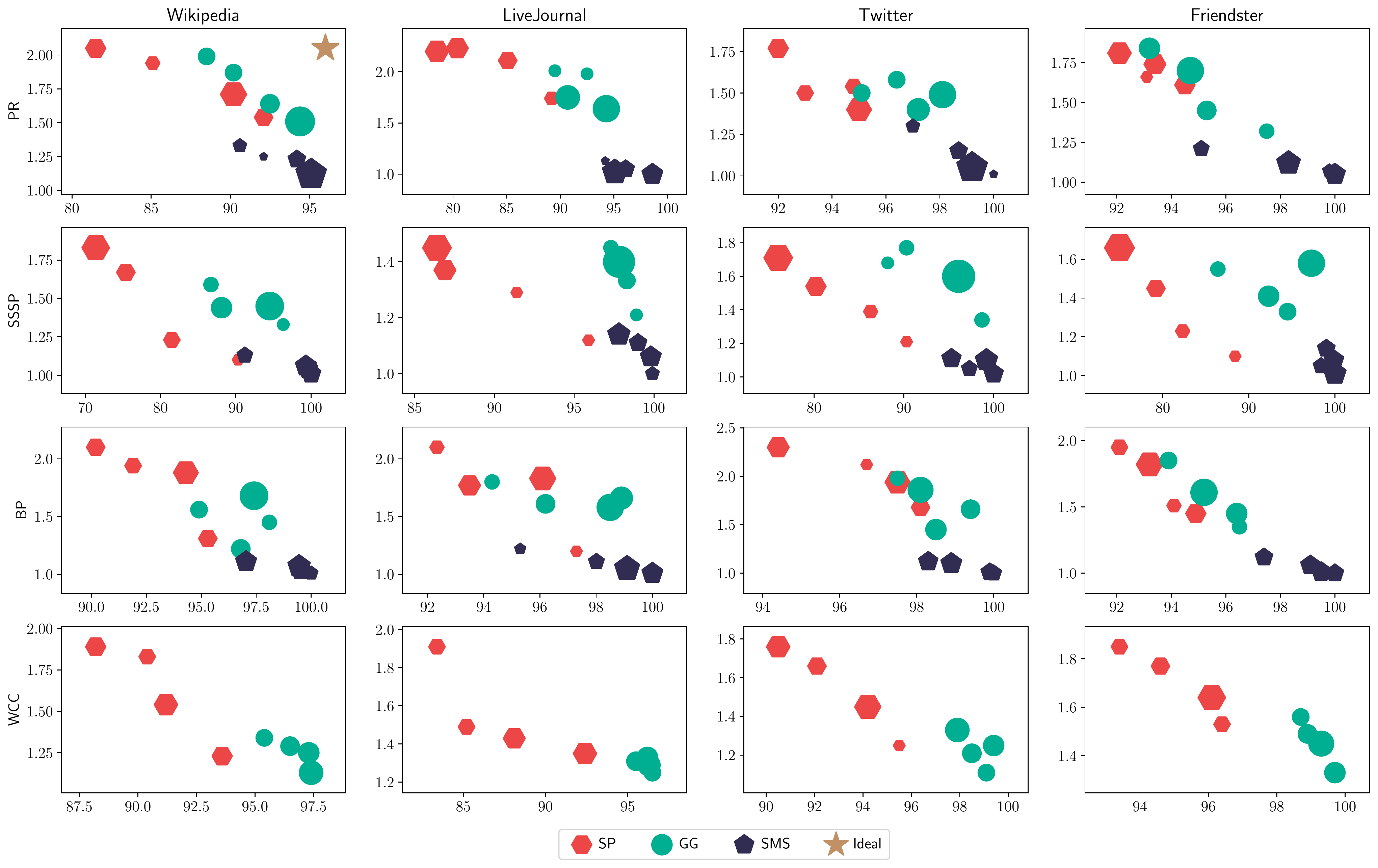}
    \vspace*{-20pt}      
    \caption{Speedup (y-axis) vs accuracy (x-axis) for three different running modes with various control parameters.
    For better visibility, we combined closer points into larger filled area.
    The ideal spot is only marked in the first figure and same for all applications/dataset.
    }
    \label{fig:pe}
    \vspace*{-10pt}
\end{figure*}

%%%%%%%%%%%%%%%%%%%%%%%%%%%%%%%%%%%%%%%%%%%%%%%%%%%%%%%%%%%%%%%%%%%%%%
\subsection{Evaluation of Performance and Accuracy}
\label{subsec:eval}

To select the best parameters for our evaluations we refer to our earlier observation in Section~\ref{subsec:sense}.
Figure~\ref{fig:agg-params} demonstrates the accumulated speedup-vs-accuracy for various configurations for PageRank on Wikipedia dataset. 
To save space, we only show PageRank in this figure, however we can see a similar pattern.
Here, the x-axis represents the accuracy and the y-axis shows the speedup over the baseline. 
We sweep through various values for each control parameter  in \GG{} ($\bm\sigma$, $\bm\theta$, and $\alpha$) and report the results as points in these figures. 
To achieve a reasonable improvement-error ratio (while eliminating the overhead of finding optimal values), we target about $90\%-95\%$ accuracy which has been shown to be an acceptable range in several previous graph approximation studies \cite{shang2014auto, heidarshenas2020v, mitliagkas2015frogwild}. 
Our experimental studies reveal that there exists a set of parameters in \GG{} that can satisfy this goal (though may not be the ideal setting for either speedup or accuracy for all applications). 
Therefore, for our experiments discussed below, we selected the parameters within the red circle for all other workloads and applications.

We compare the speedup and accuracy of \GG{} against a {\em baseline}, which is the traditional ``accurate" vertex-centric system.  
In addition, to assess the need for mode switching (between approximate vs accurate) in \GG{}, we introduce two static schemes which are special cases of \GG{}.
The first scheme is a variant of graph sparsification, referred to as \emph{SP} henceforth, where the application starts by deactivating the edges based on the parameter $\sigma$, and continues processing in this mode until the end.
Note that, unlike previous studies \cite{heidarshenas2020v, besta2019slim}, there is no need to build a new graph and we use the underlying \GG{} framework to process the graph.
The second scheme, Static Mode Switch(\emph{SMS}), uses a {\em combination} of the approximate and accurate modes. 
Similar to \emph{SP}, this scheme starts in \emph{approximate mode}, however after performing a superstep it stays in the \emph{accurate mode} for the remaining iterations.

Figure~\ref{fig:pe} compares the performance-accuracy trade-off in three aforementioned schemes (\emph{SP}, \emph{SMS}, and \emph{GG}) for all four applications and datasets. 
We use a similar setting as Figure \ref{fig:agg-params}, where the x-axis shows the accuracy and the y-axis represents the speedup. 
However, for better visibility here, we apply the kd-tree algorithm on the points for each scheme and cluster closer points into a larger filled areas. 
Note that a larger area captures the fact that more points ended up in the corresponding region. 
We also show the ideal spot (our goal), with a gold star in PR (Wiki), where the speedup and accuracy are the highest. 

In the case of PR and BP, we see that using \emph{SP} achieves a significantly higher speedup, but the accuracy is lower, due to the missing edges. 
Using \emph{SMS}, the performance improvement is limited, depending on when we switch back to the accurate mode. 
Clearly, \emph{SMS} achieves high accuracy, but there is no promising performance improvement over the baseline. 
For instance, the accuracy of \emph{SP} on \emph{Twitter} dataset is closer to that of \emph{SMS}, due to its higher density. 
Whereas, on more sparse graphs like \emph{LiveJorunal,} using \emph{SP} ends up in a lower accuracy.
On the contrary, when using \emph{GG}, we can see higher speedups with accuracy coming closer to that of \emph{SMS}. 

In SSSP, \emph{SP} performs well in terms of speedup, and its resulting accuracy is lower compared to the other two schemes. 
That is, the missing edges in \emph{SP} can exacerbate the error, since the error from one node can propagate to many other nodes. 
\emph{SMS} achieves a higher accuracy in SSSP compared to \emph{SP}, as expected; however, the accuracy can vary depending on the start of the accurate mode. 
This also can hinder the performance of the \emph{SMS} scheme. 
In comparison, \emph{GG} brings the best of both worlds, and helps us achieve an accuracy which is very similar to that of \emph{SMS}, and a performance which is very close to that of \emph{SP}.

% \setlength{\tabcolsep}{0.15em}
% \begin{wrapfigure}[8]{R}{0.6\textwidth}
% \begin{minipage}{0.55\textwidth}
% \vspace{-40pt}
\begin{table}[H]
\caption{Comparison of speedup (Spd) and accuracy (Acc) between GG and others}
\label{table:compare}
\vspace{-5pt}
\centering
\resizebox{0.95\linewidth}{!}{
\begin{tabular}{@{}ccccccccc@{}} \toprule
\multicolumn{2}{c}{\textbf{Alg (Dataset)}}  & \textbf{PR (LJ)} & \textbf{PR (TW)} & \textbf{PR (FS)} & \textbf{BP (LV)} & \textbf{BP (TW)} & \textbf{BP (FS)} & AVG \\ \midrule
\multirow{2}{*}{\texttt{GG}}    & Spd ($\times$)    & $1.64$  & $1.49$    & $1.66$    & $1.68$  & $1.85$    & $1.62$  & $1.66$    \\ 
                                & Acc ($\%$)        & $94.32$ & $98.12$   & $94.74$   & $96.69$ & $98.08$   & $95.29$ & $96.20$   \\ \midrule
                                %%%%%%%%%%%%%%%%%%%%%%%%%%%%%%%%%%%%%%%%%%%%%%%%%%%%%%%%%%%%%%%%%%%%%%%%%%%%%%%%%%%%%%%%%%%%%%%%%%%%%%%%%%%%%%%%%%%%%%%%%%%%%%
\multirow{2}{*}{\texttt{SP}}    & Spd ($\times$)    & $1.74$  & $1.54$    & $1.81$    & $1.83$  & $1.94$    & $1.81$  & $1.78$    \\ 
                                & Acc ($\%$)        & $89.21$ & $94.84$   & $92.19$   & $93.13$ & $97.57$   & $93.64$ & $93.43$   \\ \midrule
                                %%%%%%%%%%%%%%%%%%%%%%%%%%%%%%%%%%%%%%%%%%%%%%%%%%%%%%%%%%%%%%%%%%%%%%%%%%%%%%%%%%%%%%%%%%%%%%%%%%%%%%%%%%%%%%%%%%%%%%%%%%%%%%
\multirow{2}{*}{\texttt{VC}}    & Spd ($\times$)    & $1.45$  & $1.34$    & $1.12$    & $1.20$  & $1.16$    & $1.29$  & $1.26$    \\ 
                                & Acc ($\%$)        & $87.15$ & $95.36$   & $91.63$   & $94.41$ & $97.91$   & $93.41$ & $93.31$   \\
\bottomrule
\end{tabular}
}
\end{table}
% \end{minipage}
% \end{wrapfigure}

In WCC, the vertex property is defined as the \emph{Connected Component ID} of the vertex residing in. 
Consequently, the influence estimate is a binary decision -- whether the ID is changed or not. 
This implementation of WCC application forces the influence values to be either $0$ or $1$. 
Hence \emph{GG} and \emph{SMS} end up exhibiting the same behavior; so, we only show \emph{GG} in this figure. 
As can be observed in Figure \ref{fig:pe}, \emph{GG} performs much better (as far as error is concerned) compared to \emph{SP}, with a lower speedup. 
This also proves that \emph{GG} is more flexible in terms of application, and can be applied in various settings depending on the need.

{\textbf{Overall Performance and Accuracy:}}
We compare the speedup and accuracy values achieved by \GG{} against a recent approximate graph processing system, V-Combiner~\cite{heidarshenas2020v}. While in this paper we are aiming at minimal pre-processing, we include V-Combiner due to its lower overhead compared to the alternative methods. Note also that V-Combiner only supports specific types of graph applications such as PR and BP, while \GG{} can support any graph algorithm implemented on current graph processing systems. We also include sparsification (\emph{SP}) as a special case of \GG{} without adaptive correction, as discussed earlier in Section~\ref{subsec:eval}.

Table~\ref{table:compare} shows the speedup and accuracy for \GG{}, Sparsification, and V-Combiner, on PR and BP, using three different datasets on top 10 best configurations. It is to be noted that V-Combiner does not support SSSP and WCC. From this experiment, we observe that Sparsification has the best speedup, close to $72\%$ on average, at the cost of lower accuracy. However, V-Combiner suffers from lower speedup due to mandatory pre-processing, which involves additional graphs creation and the recovery phase. The performance gain of V-combiner on average is about $26\%$ compared to the baseline, while it maintains an acceptable accuracy. Finally, our proposed approach, \GG{}, closes the gap between these two with a speedup of up to $85\%$ and $58\%$ on average, while maintaining a higher accuracy compared to the other two methods.
%%%%%

\section{Acknowledgements}
This work was supported in part by CRISP, one of six centers in JUMP, a Semiconductor Research Corporation (SRC) program sponsored by DARPA and NSF grants 1909004, 1714389, 1912495, 1629915, 1629129, 1763681.

\section{Concluding Remarks}
\label{section:conclusion}

This paper presents \GG{}, a novel attempt at approximating graph processing with a simple extension to current APIs. Inspired by the ideas from graph theory and approximation analysis, \GG{} implements an adaptive approximate graph processing strategy that requires no time-consuming preprocessing and can be applied to any graph processing system.
\GG{} preserves the main characteristics of the graph using adaptive correction and provides sufficient flexibility in modulating different control parameters. In this work, we vary these control parameters to achieve performance-accuracy trade-offs, and our experimental studies show that \GG{} achieves up to $1.85\times$ speedup while maintaining high accuracy compared to an accurate baseline. 
% In our future work, we intend to explore techniques for automatically selecting and tuning these control parameters based on application characteristics. 

\bibliographystyle{splncs04}
\bibliography{reference}

\end{document}